\documentclass[12pt]{article}
\usepackage[dvips]{color}
\usepackage{amsmath}
\usepackage{graphicx}
\usepackage{subfigure}
\usepackage{epsfig}
\usepackage{amsmath}
\usepackage{cite}
\usepackage{color}
\usepackage{subfigure}

\begin{document}
\begin{center}
\Large{\bf Inflation, Swampland,  and Landscape}\\
\small \vspace{1cm}
{\bf S. Noori Gashti$^{\star,\dagger}$\footnote {Email:~~~saeed.noorigashti@stu.umz.ac.ir}}, \quad
{\bf J. Sadeghi$^{\star}$\footnote {Email:~~~pouriya@ipm.ir}}, \quad
\\
\vspace{0.5cm}$^{\star}${Department of Physics, Faculty of Basic
Sciences,\\
University of Mazandaran
P. O. Box 47416-95447, Babolsar, Iran}\\
$^{\dagger}${School of Physics, Damghan University, P. O. Box 3671641167, Damghan, Iran
\small \vspace{1cm}
}
\end{center}
\begin{abstract}
In this paper, we evaluate a new refined de Sitter (dS) conjecture perspective. This conjecture provides interesting conditions in studying various inflationary models.  Therefore we challenge refined dS conjecture with a general method for analyzing the potentials with the inflectional point near the top ($V''=0$). We find the compatible spaces for our inflationary model that live out of swampland according to the latest observational data, such as Planck 2018.\\
Keywords: Inflationary Model, Refined dS Conjecture, Swampland
\end{abstract}
\newpage
\tableofcontents
\section{Introduction}

Recently,  researchers solved some of the cosmological problems  \cite{1,2}, but with the advancement of science, new questions always arise. For example, how can one describe UV complete theories compatible with quantum gravity which is called the landscape?  Another concept that is vaster than the landscape is called swampland, inconsistent with quantum gravity. Many authors studied various models to determine which of these effective field theories can be in landscape (consistent with quantum gravity) or the swampland\cite{3}. Also, researchers have considered the Swampland program, i.e., the weak gravity conjecture (WGC) and other conditions such as swampland, landscape, TCC, and studied them under different conditions and theories viz physics of black holes, inflation, dark energy, etc.
The ones analyzed various cosmological concepts and compared them with the latest observable data. To more understand about these conditions and their implications, you can see in Ref.s \cite{3,4,5,6,7,8,9,10,11,12,13,14,15,16,17,18,19,20,21,22,23,24,25,26,27,28,29}. In recent years, Ooguri, Palti, Shiu, and Vafa proposed the refined swampland dS conjecture \cite{30}. Already the older version of this conjecture is suggested in\cite{31}. This conjecture suggests that all the scalar potential for any effective field theory that is compatible with string theory, in other words, consistent with quantum gravity, must be satisfied with one of the following two conditions, which are proportional to the derivatives of the potential. These conditions are expressed in the following form.

\begin{equation}\label{1}
|\nabla V|\geq c_{1} \frac{V}{M_{pl}} \hspace{70pt}
 \end{equation}

\begin{equation}\label{2}
min(\nabla_{i}\nabla_{j}V)\leq -c_{2}\frac{V}{M_{pl}^{2}} \hspace{20pt}
 \end{equation}

Where $c_{1}$ and $c_{2}$ are the constant parameters and $(min(\nabla_{i}\nabla_{j}V))$ is the minimum eigenvalue of $(\nabla_{i}\nabla_{j}V)$. With consideration of the above conjecture, we can establish one of the most straightforward implications of them about cosmological constants, for $|\nabla V_{c.c.}|=0$ and $ V_{c.c.}>0 $, is ruled out. \cite{32,33,34,35,36,37,38,39,40,41,42,43,44,45,46,47,48,49,50,51,52,67} With respect to above conditions, it is noteworthy that the second condition can also be easily satisfied for any potential in  $\Delta\phi\ll M_{pl}$ scale.

\begin{equation}\label{3}
M_{pl}^{2}\frac{\nabla_{i}\nabla_{j}V}{V}\sim -\frac{M_{pl}^{2}}{\Delta \phi^{2}}\ll -c_{2}
 \end{equation}

Inflationary dynamics usually occur at high scales and discussed their limitations in \cite{33}. The swampland programs had interesting results in the literature.
So the mentioned concept is the motivation for us to evaluate different cosmological concepts such as inflationary models from the dS conjecture perspective. In this paper, we will introduce a specific inflationary model and express its characteristics. We challenge this model with refined swampland conjecture and describe its implications. It is possible to determine the areas where this potential is located out of the swampland. With respect to all of the above concepts, we have organized this paper as follows.\\
In section 2, we introduce the inflationary model (Tip Inflation ) . Then we apply the refined swampland conjecture to it. We describe the results by plotting some figures and determining the compatible areas with the swampland conjectures under the model's free parameters in section3, and we describe the results in section4.
\section{Tip Inflation}

This inflationary model is a typical theory based on string theory, which can interpret as the branes motion, which means these motions In the extra dimensions cause the 4D space-time to swell. The theoretical justification of tip inflation model investigate with respect to a K$\ddot{a}$hler potential, KKLT, the Klebanov-Strassler throat, etc., So in this route, ones assume a D3-brane moving at the tip, a module structure and dilaton due to the fluxes presence are stabilized and according too\cite{64} there is only one volume module. Also, three fields $z_{i}, i=1, 2, 3$ that describe the D3-brane state. In that case, it follows a form of K$\ddot{a}$hler potential.

\begin{equation}\label{4}
K\bigg(\rho, z_{i},z^{\dag}_{i}\bigg)=-3M_{pl}^{2}\ln\bigg[\rho+\rho^{\dag}-\gamma k(z_{i},z^{\dag}_{i})\bigg]
 \end{equation}

Where $k$ and $\gamma$ are a brane coordinates function and a constant, respectively. An approximate expression obtains Proportional to the $T_{3}$ brane tension. Also, in the deformed conifold tip neighborhood, the function $k$ is expressed as follows.

\begin{equation}\label{5}
k(z_{i},z^{\dag}_{i})=k_{0}+c\varepsilon^{-3/2}\bigg(\Sigma_{\mathcal{A}=1}^{4}|Z_{\mathcal{A}}|^{2}-\varepsilon^{2}\bigg)
 \end{equation}

Where $c=2^{1/6}/3^{1/3}\approx0.77$ is a constant and $k_{0}$ is also dependent on the values of $k$ at the tip. Also, another important element of the model is an n D7-branes placed far from the tip. Then superpotential\cite{65} It is expressed in the following form.

\begin{equation}\label{6}
\mathcal{W}=\mathcal{W}_{0}+\mathcal{A}(z_{1})\exp(-a\rho)=\mathcal{W}_{0}+\mathcal{A}_{0}\bigg(1-\frac{z_{1}}{\mu}\bigg)^{1/n}\exp(-a\rho)
 \end{equation}

In the above equation $\mathcal{W}_{0}$, $\mathcal{A}_{0}$ and $a$ are constant parameters. We will also explain the parameter $\mu$ in detail below. The great thing here is that the above superpotential only depends on $z_{1}$ hence breaking the symmetry of the tip. in the following, we need to introduce an important potential which is called the F-term potential,

\begin{equation}\label{7}
\begin{split}
V(\sigma, x_{1})=&\frac{2a\exp(-a\sigma)}{M_{pl}^{2}U^{2}}\bigg(\frac{aU}{6}|\mathcal{A}|^{2}\exp(-a\sigma)+|\mathcal{A}|^{2}\exp(-a\sigma)-|\mathcal{W}_{0}\mathcal{A}|\bigg)\\
&+\frac{\exp(-2a\sigma)|\mathcal{A}|^{2}\varepsilon^{2/3}}{3M_{pl}^{2}\gamma U^{2}n^{2}\mu^{2}c}\bigg(1-\frac{x_{1}^{2}}{\varepsilon^{2}}\bigg)\bigg(1-\frac{x_{1}}{\mu}\bigg)^{-2}+\frac{D}{U^{b}}
\end{split}
 \end{equation}

It can also be described as $z_{1}=x_{i}+iy_{i}$ and $z_{1}=x_{1}$ at the tip. $V$ don't depend on $x_{2}$ and $x_{3}$ corresponding to our superpotential selection.
We can also define $\rho=\sigma+i\tau$ and $U=\rho+\rho^{\dag}-k=2\sigma-k_{0}$. The uplifting term in the final sentence has two constant parameters, D and b, which was added because it avoided the anti-de Sitter minimum. Calculating a kinetic sentence with this method is complicated because the K$\ddot{a}$hler matrix mixes all the fields $z_{i}$.
So we should introduce another structure as $$z_{1}=\varepsilon\cos\varphi$$
$$z_{2}=\varepsilon\sin\varphi\cos\theta$$
$$z_{3}=\varepsilon\sin\varphi\sin\theta\cos\psi$$
$$z_{4}=\varepsilon\sin\varphi\sin\theta\sin\psi $$
that creates a good result, i.e., the K$\ddot{a}$hler matrix becomes diagonal and expanding everything with respect to the small parameter $\varepsilon/\mu\ll 1$. Hence we will have.

\begin{equation}\label{8}
V(\sigma, \varphi)=\Lambda(\sigma)+B(\sigma)\cos\varphi+C(\sigma)\sin^{2}\varphi+...
 \end{equation}

According to the above equation,

\begin{equation}\label{9}
\begin{split}
&\Lambda(\sigma)=\frac{2a|\mathcal{A}_{0}|\exp(-a\sigma)}{M_{pl}^{2}U^{2}}\times\bigg(\frac{aU}{6}|\mathcal{A}_{0}|\exp(-a\sigma)+|\mathcal{A}_{0}|\exp(-a\sigma)-|\mathcal{W}_{0}|\bigg)+\frac{D}{U^{b}}\\
&B(\sigma)=\frac{2a|\mathcal{A}_{0}|\exp(-a\sigma)\varepsilon}{M_{pl}^{2}U^{2}n\mu}\times\bigg(-\frac{aU}{3}|\mathcal{A}_{0}|\exp(-a\sigma)-2|\mathcal{A}_{0}|\exp(-a\sigma)+|\mathcal{W}_{0}|\bigg)\\
&C(\sigma)=\frac{|\mathcal{A}_{0}^{2}|\exp(-a\sigma)\varepsilon^{2/3}}{3M_{pl}^{2}U^{2}n^{2}\mu^{2}\gamma c}\\
\end{split}
 \end{equation}

With looking carefully at the above equations, if we ignore all the sentences related to the brane state, then only $\Lambda(\sigma)$ remains, which is nothing but the (KKLT) potential for the volume modulus \cite{66}.
Also, by removing the important term $D/U^{b}$, $\partial\Lambda/\partial\sigma$=0 investigated the minimum which located at $\sigma=\sigma_{0}$, and the tacit solution which is given by

\begin{equation}\label{10}
\begin{split}
\mathcal{W}_{0}=-\mathcal{A}_{0}\bigg(1+\frac{a}{3}(2\sigma_{0}-k_{0})\bigg)\exp(-a\sigma_{0})
\end{split}
 \end{equation}

The corresponding quantity of potential be negative viz (anti de Sitter), So

\begin{equation}\label{11}
\begin{split}
\Lambda(\sigma_{0})=\frac{a^{2}\mathcal{A}_{0}^{2}}{3M_{pl}^{2}U}\exp(-2a\sigma_{0})<0
\end{split}
 \end{equation}

hence the existence of an uplifting term creates a new minimum at V that is positive.
This issue shows that how KKLT organizes a de Sitter minimum instead of an anti-de Sitter, which was first studied in string theory \cite{66}.
If the minimum situation did not change by adding the mentioned sentence viz uplifting term $D/U^{b}$, one would calculate a vanishing V for

\begin{equation}\label{12}
\begin{split}
D_{0}=\frac{a^{2}\mathcal{A}_{0}^{2}U^{b-1}(\sigma_{0})}{3M_{pl}^{2}}\exp(-2a\sigma_{0})
\end{split}
 \end{equation}

that led to the definition of a new parameter $\beta$ as follows.

\begin{equation}\label{13}
\begin{split}
\beta\equiv D\times\frac{3M_{pl}^{2}}{a^{2}\mathcal{A}_{0}^{2}U^{b-1}(\sigma_{0})}\exp(2a\sigma_{0})
\end{split}
 \end{equation}

In general, The correction can evaluate due to the uplifting terms and can be obtained in the following form.

\begin{equation}\label{14}
\begin{split}
\sigma_{min}=\sigma_{0}+\frac{b\beta}{2a^{2}\sigma_{0}}+...
\end{split}
 \end{equation}

Where provided a valid form as $b\beta/(2a^{2}\sigma_{0})\ll1$. Also, by electing the $\beta=0$ , we will have $\sigma_{min}=\sigma_{0}$.
the interesting point is that some other corrections are also possible for the position of the minimum due to the presence of the brane, but one can determine that none of them play an important role \cite{64}.
Considering that the modulus is stabilized at this minimum; hence we present a single-field model $V(\varphi)=V(\sigma_{min}, \varphi)$ with respect to the components of the equation (8) in the following form.

\begin{equation}\label{15}
\begin{split}
&\Lambda(\sigma_{min})\equiv\Lambda\simeq\frac{a^{2}|\mathcal{A}_{0}^{2}|\exp(-2a\sigma_{0})}{6M_{pl}^{2}\sigma_{0}}\bigg[(\beta-1)+...\bigg]\\
&B(\sigma_{min})\equiv B\simeq\frac{a|\mathcal{A}_{0}^{2}|\varepsilon\exp(-2a\sigma_{0})}{6M_{pl}^{2}\sigma_{0}^{2}n\mu}\bigg[(b\beta-3)+\frac{b\beta}{4a\sigma_{0}}(14-3b\beta)+...\bigg]\\
&C(\sigma_{min})\equiv C\simeq\frac{|\mathcal{A}_{0}^{2}|\varepsilon^{2/3}\exp(-2a\sigma_{0})}{12M_{pl}^{2}\sigma_{0}^{2}n^{2}\mu^{2}\gamma c}+...
\end{split}
\end{equation}

The above equations show the potential parameters in terms of string parameters. As we can see, for $\beta>1$, we have a potential (KKLT) that is positive at minimum that could account for a cosmological constant for $\beta-1=\mathcal{O}(\sigma_{0}^{_2})$\cite{64}. Using a clear form, we will have K$\ddot{a}$hler metric.

\begin{equation}\label{16}
\begin{split}
K_{I\overline{J}}\partial_{\mu}z^{I}\partial^{\mu}z^{\overline{J}}\simeq\frac{3M_{pl}^{2}}{U}\gamma c \varepsilon^{4/3}\partial_{\mu}\varphi\partial^{\mu}\varphi
\end{split}
 \end{equation}

Which we will, at minimum,

\begin{equation}\label{17}
\begin{split}
\gamma\simeq\frac{\sigma_{0}T_{3}}{3M_{pl}^{2}}
\end{split}
 \end{equation}

Where $T_{3}$ indicates the brane tension. According to the above concepts in the large volume limit, the canonical field $\phi$ will be in the form $\phi=\sqrt{T_{3}c}\varepsilon^{2/3}\varphi$. As a valuable result, the final general form of this potential is mentioned as follows,\cite{53}

\begin{equation}\label{18}
V(\phi)=\Lambda+B\cos(\frac{\phi}{\sqrt{T_{3}c}\varepsilon^{\frac{2}{3}}})+C\sin^{2}(\frac{\phi}{\sqrt{T_{3}c}\varepsilon^{\frac{2}{3}}})
 \end{equation}

It would be interesting to discuss the magnitude orders of the parameters displayed in the above potential. So, parameter $\sigma_{0}$ is the volume modulus related to the size or volume of extra dimensions, $V_{6}\simeq \sigma_{0}^{3/2}\alpha'^{3}$. Then you can define brane tension in form $T_{3}=(2\pi)^{-3}g_{s}^{-1}\alpha'^{-2}$, and Planck mass takes as $M_{pl}^{2}=2(2\pi)^{-7}V_{6}g_{s}^{-2}\alpha'^{-4}$. In these equations, $g_{s}$ is called string coupling. The distance $\mu^{2/3}$ can also see as the distance between the stack of D7-branes and the tip. Also, about the size of the throat, which can state this parameter as $\mu\simeq(27\pi g_{s}\mathcal{N}\alpha'^{2}/4)^{3/8}$. Among the other parameters under consideration $\mathcal{N}$, which refers to as total background Ramond-Ramond charge. Given the concepts mentioned, we must assume that the potential disappears at its minimum value to have a suitable slow-roll scenario. we will have $\beta_{sr}=1+\frac{45\varepsilon}{4n\mu a^{2}\sigma_{0}^{2}}$ by considering the $\Lambda=B$, also with respect to  $\beta=\beta_{sr}$ with $b=3$. The above concept showed that we performed a large volume expansion. Therefore
at the top of The potential has a relation as $\partial^{2}V/\partial\phi^{2}\simeq 2C-\Lambda$, and if it wants a flat potential $2C-\Lambda=2C-B$, it must be very small values. That's mean $C/B\simeq1/2$. Using the above concepts can be written.
\begin{equation}\label{19}
\frac{C}{B}=\Upsilon\times\frac{\sigma_{0}^{3/2}}{g_{s}(g_{s}\pi\mathcal{N})^{3/8}}(\frac{r_{tip}}{\ell_{s}})^{-1/2}
 \end{equation}
According to the above equation, we have $\Upsilon=(12/15)\times(4/27)^{3/8}/((2\pi)^{4}nc)\approx5\times 10^{-5}$ and $r_{tip}\equiv \varepsilon^{2/3}$. String length is also showed by $\ell_{s}=\sqrt{\alpha'}$. Another interesting point is the discussion of mass scale, which usually appears in discussing arguments of trigonometric functions. We also have.
\begin{equation}\label{20}
\frac{\sqrt{T_{3}c}\varepsilon^{2/3}}{M_{pl}}=(2\pi)^{2}\sqrt{\frac{c}{2}}g_{s}^{1/2}\sigma_{0}^{-3/4}(\frac{r_{tip}}{\ell_{s}})
 \end{equation}

the radius of the tip and the volume of the extra dimensions control two inflation parameters $C/B$ and $\sqrt{T_{3}c}\varepsilon^{2/3}/M_{pl}$ according to the above concepts for constant parameters $g_{s}$ and $\mathcal{N}$. , the above equation implies that $\sqrt{T_{3}c}\varepsilon^{2/3}/M_{pl}\simeq2\times 10^{8}\sigma_{0}^{9/4}$ when $C/B=1/2$, like the approximation used in the slow-roll analysis. Therefore, We will have an equation in the following form for the canonically normalized inflaton field according to the above concepts and the slow-roll analysis of the mentioned model and with respect to inflation progress in the  $0<\phi/\mu<\pi$.

\begin{equation}\label{21}
V=M^{4}\bigg(1+\cos(\frac{\phi}{\mu})+\alpha \sin^{2}(\frac{\phi}{\mu})\bigg)
 \end{equation}

Here we rewrite $\Lambda=M^{4}$, $C/B=\alpha$ and $\mu=\sqrt{T_{3}c}\varepsilon^{\frac{2}{3}}$. This potential converts to natural inflation (NI) by considering $\alpha\ll 1$. Also, this model does not appear in $\alpha\ll 1$ . in $\alpha\simeq 1/2$, the model becomes very flat at the top, leading to a phenomenologically successful slow-roll inflationary stage could occur. We investigate potentials with uniform growth and typically inflectional points near the top in ($V''=0$). In this paper, we examine the Tip inflationary model according to the above concepts. Also, we consider $\alpha=1/2$ and $\mu/M_{pl}=\mathcal{O}(10^{-4})$ to establish the slow-roll conditions proposed for this model. As can see in the below figures, we plot the potential in different ranges; $(0,\pi)$ and $(0,\frac{\pi}{3})$. Inflation starts almost below the peak, such as $\frac{\phi}{\mu}\approx \frac{\pi}{3}$ and then roll down to the true vacuum. As mentioned above, our potential at certain level decreases to natural inflation (NI). The shape of this potential is determined in figure (1).

\begin{figure}[h!]
 \begin{center}
 \subfigure[]{
 \includegraphics[height=5cm,width=5cm]{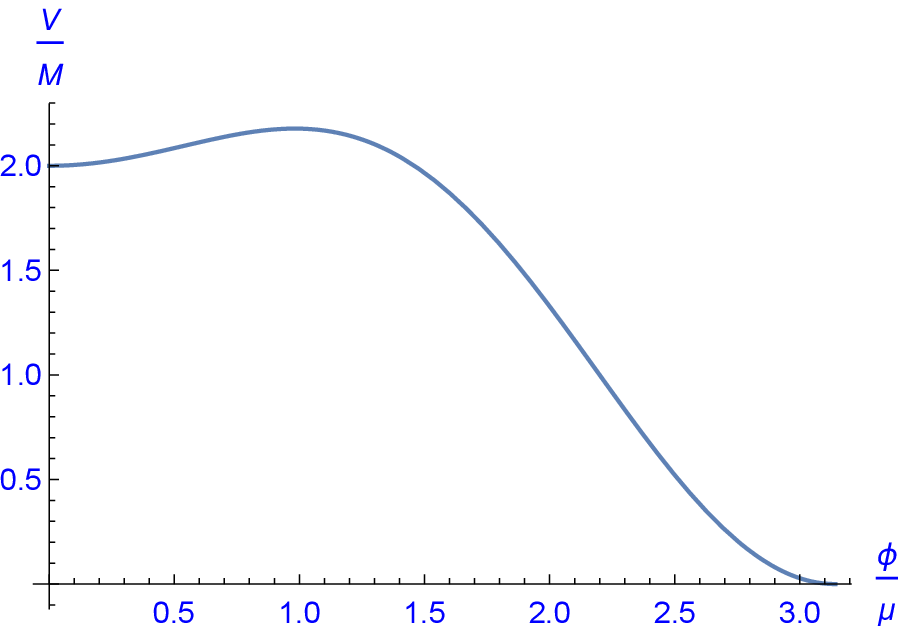}
 \label{1a}}
 \subfigure[]{
 \includegraphics[height=5cm,width=5cm]{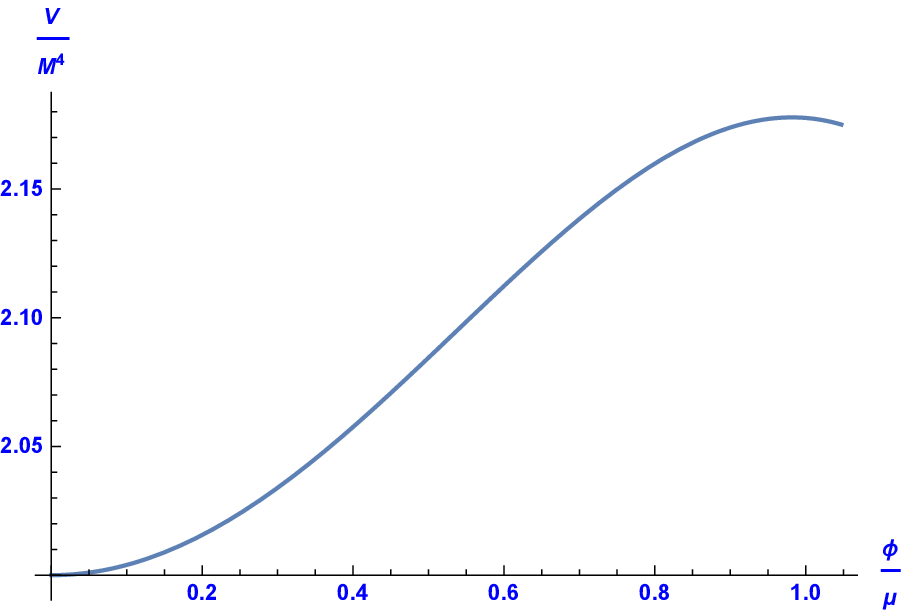}
 \label{1b}}
  \caption{\small{ The potential for the Tip inflation model in $\frac{\phi}{\mu}= (0,\pi)$ and $(0,\frac{\pi}{3})$ in (1a) and (1b) respectively. }}
 \label{4}
 \end{center}
 \end{figure}
\section{Tip Inflation $\&$ RSC}
\subsection{Case I}

We benefit from the  mentioned earlier concept, and we use the first condition viz according to equation (1), we define a function in the following form,

\begin{equation}\label{22}
F(\frac{\phi}{\mu})\equiv \mu\frac{||V'||}{V}=\frac{(-1+2\alpha \cos(\frac{\phi}{\mu}))\sin (\frac{\phi}{\mu})}{1+\cos(\frac{\phi}{\mu})+\alpha \sin^{2}(\frac{\phi}{\mu})}
 \end{equation}

According to figure (2), We know that case (1) establishes a convention in the upper limit for $\phi$.

\begin{equation}\label{23}
F(\frac{\phi}{\mu})\geq \mu\frac{||V'||}{V} \Leftrightarrow \phi < \phi_{\bullet}
 \end{equation}

Critical values $\phi_{\bullet}$ obtained with respect to $c_{1}\frac{\mu}{M_{p}}$, which can accurately calculate according to observable data such as Planck 2018.

\subsection{Case II}

We can determine the inflection point with ($ V ''= 0 $).  We introduce another contractual function, (G), containing information about potential curvature to analyze the second condition. So this function is defined as follows,

\begin{equation}\label{24}
G(\frac{\phi}{\mu})\equiv \mu^{2}\frac{||V''||}{V}=\frac{-\cos(\frac{\phi}{\mu})+2\alpha \cos(2\frac{\phi}{\mu})}{1+\cos(\frac{\phi}{\mu})+\alpha \sin^{2}(\frac{\phi}{\mu})}
 \end{equation}

We also know that case (2) establishes a new contract in lower limit for $\phi$ like the previous part, and according to figure (2),

\begin{equation}\label{25}
G(\frac{\phi}{\mu})\leq -c_{2}\frac{\mu^{2}}{M_{pl}^{2}}\Leftrightarrow \phi > \phi_{\ddag}
 \end{equation}

Where we can obtain the $$\phi_{\ddagger}$$ from $c_{2}\frac{\mu}{M_{p}}$.
critical values of $\phi$ can investigate for case (1) and case (2) by defining two contract functions containing curves' information with respect to the above conditions and the latest observable data,
These values are well represented in figure (2). The location of two critical values, $(\phi_{\bullet})$ (Intersection point between ($blue$) and ($green$), and $(\phi_{\ddagger})$ (Intersection point between $(red)$ and $(orange)$ are specified in this figure.

\begin{figure}[h!]
 \begin{center}
 \subfigure[]{
 \includegraphics[height=5cm,width=5cm]{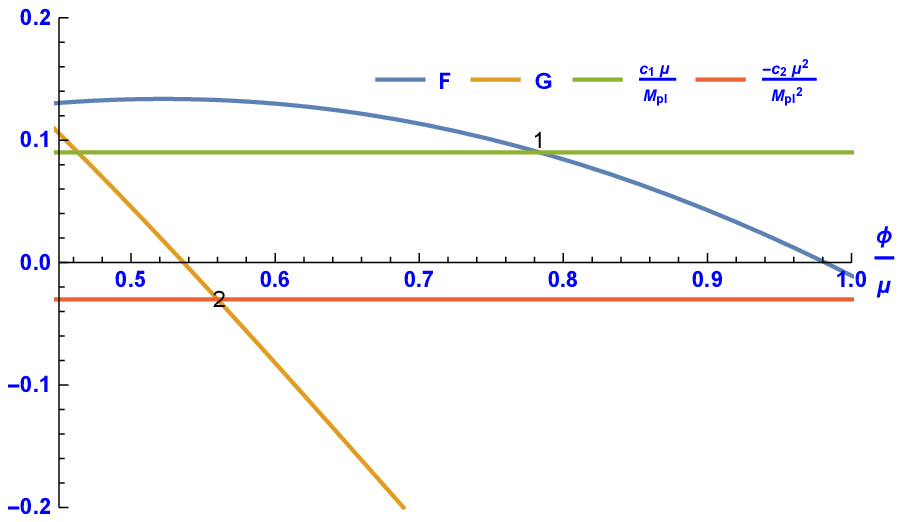}
 \label{2a}}
 \caption{\small{$F(\frac{\phi}{\mu})$ and $ G(\frac{\phi}{\mu})$ in $(\frac{\phi}{\mu})$. The location $1=(\phi_{\bullet})$ (Intersection point between ($blue$) and $green$), and $2=(\phi_{\ddagger})$ (Intersection point between (red) and (orange))}}
 \label{2}
 \end{center}
 \end{figure}

\newpage
\subsection{Unification of Case I and Case II}

In this subsection, we examine the combination of the case (I) and the case (II) with respect to the concepts discussed in the previous two-part.  The first condition in equations (22, 23) and the second condition in equations (24, 25) are completely independent. However, at least one of these samples can be satisfying in all regions of $\phi$; provided that the lower bound the second case $\phi_{\ddagger}$ is smaller than the upper bound of the first case $\phi_{\bullet}$, or in other words, one can obtain the following condition.

\begin{equation}\label{26}
\frac{\phi_{\ddagger}}{\mu}\leq \frac{\phi_{\bullet}}{\mu}
 \end{equation}

The inequality in the equation (26) establishes a general condition for an inflationary model with a specific feature. The common feature is the potential that has an inflection point upwards. There are many models of this type, such as Higgs inflation\cite{54,55,56,57,58,59}. As shown in figure (2), $(F\geq G)$ in $(\frac{\phi}{\mu})\geq0.45$, so that a set of values $(c_{1},c_{2})$, can be obtained for given value $\frac{\mu}{M_{pl}}$ that satisfies the desired condition, i.e., the swampland dS criteria are satisfied. The allowable range is specified in the figure with respect to various values of $\frac{\mu}{M_{pl}}$. for instance, for the approximate of $(c_{1},c_{2})\approx(0.09,0.04)$ and $\frac{\mu}{M_{pl}}\approx 0.85$ the value of $(\phi_{\ddagger},\phi_{\bullet})\approx(0.55\mu,0.8\mu)$ which is proportional to the swampland ds conjecture. This condition reviewed for other values of constant parameters. Also in figure (3), we have plotted critical lines according to $(c_{1},c_{2})$, for different values of $(\frac{\mu}{M_{pl}})$.
The critical values of $(\phi_{\ddagger}\leq \phi_{\bullet})$, are satisfied the below areas of the lines, viz the refined swampland dS conjecture has been met and be in Landscape. Our potential may not be compatible with the quantum gravitational UV completion at areas above the lines. It may belong to an area that is incompatible with quantum gravity or swampland. Also, the reheating compatible with slow-roll predictions of the mentioned model have been performed for different values of the free parameters and have led to exciting results, such as for $\alpha>1/2$ and $\alpha<1/2$ with respect to different values of $\mu/M_{pl}$\cite{53}. In both examples, should set it to $\alpha=1/2$, or $|2\alpha -1|\ll \mu^{2}/M_{pl}^{2}$. Otherwise, there is a very important point which is deviating from scale invariance. The typical values of gravitational waves are very small.  Also, in this model, ones attempt to constrain the value of $\mu/M_{pl}$ in return for its slow-roll predictions and $\alpha=1/2$. If one considers the values of $\mu/M_{pl}$ larger than the $10^{-4}$, these models will not be interesting by observations since they deviate too much from scale invariance. We also have considered different quantities in our calculations; the allowable range for swampland conjectures is well defined, and a kind of compatibility with the above concepts is clear. It is an interesting result that our model is completely in Landscape for the amount of $\mu/M_{pl}$ mentioned.

\begin{figure}[h!]
 \begin{center}
 \subfigure[]{
 \includegraphics[height=5cm,width=5cm]{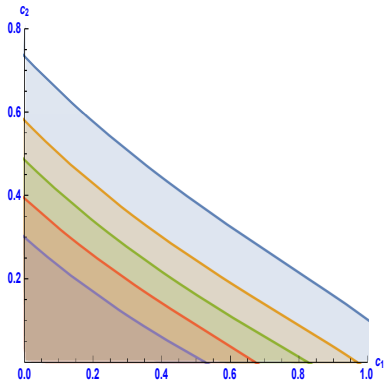}
 \label{3a}}
 \caption{Critical lines $(c_{1}, c_{2})$ with respect to various values of $(\frac{\mu}{M_{pl}})$. The areas above the lines are not excluded by dS conjecture. In fact, it can be said that it belongs to the swampland. The lines from up to down $(\frac{\mu}{M_{pl}})=0.85, 0.95, 1.05, 1.15, 1.25$ }
 \label{3}
 \end{center}
 \end{figure}

This paper investigated the implications of refined dS conjecture for the specific inflationary model (Tip inflation). By using a series of limitations  and observable data such as (Planck+BICEP2+Keck)\cite{60,61,62,63} were determined the allowable range for mentioned model with respect to universal values $(c_{1},c_{2})$. These universal values can be limited under certain conditions, but it is clear and definite that $(c_{1}\leq 1,c_{2}\leq 1)$. However, the above conditions can consider any potential with uniform growth and an inflection point near the top.
\section{Conclusions}
Recently, most cosmologists have paid attention to various inflationary models with respect to different conditions such as slow-roll, constant roll, etc.,  to study their implications to understand the universe's structure. In this paper, we also evaluated the new perspective of the refined ds conjecture. This conjecture provides interesting points for a variety of inflationary models. We have considered a general method for analyzing potential with an inflection near top points ($ V '' = 0 $). We've found compatible parametric spaces where the inflationary model lives in the landscape instead of being in swampland based on the latest observational data, such as Planck 2018. We chose an inflation model based on the string theory viz tip inflationary model and applied the refined swampland conjectures. We determined the permissible areas with swampland conjectures by putting two conditions according to the free parameters $\alpha$ and $\mu/M_{pl}$, also by obtaining the critical lines with respect to $(\phi_{\ddagger}$, $\phi_{\bullet})$, and the latest observable data. This assumption can apply to different types of inflationary models. It will be interesting to categorize the inflation models according to the limitations associated with the swampland program.


\begin{thebibliography}{11}
\bibitem{1}
C. Vafa, hep-th/0509212 (2005).
\bibitem{2}
H. Ooguri and C. Vafa, Nucl. Phys. B766 21–33, (2007).
\bibitem{3}
E. Palti, Fortsch. Phys. 67 6, 1900037 (2019).
\bibitem{4}
N. A. Hamed, L. Motl and A. Nicolis, JHEP 0706, 060 (2007).
\bibitem{5}
A. Medved, Mod. Phys. Lett. A, 22, 2605-2610 (2007).
\bibitem{6}
A. Urbano, arxiv: 1810.05621 (2018).
\bibitem{7}
 Y. Akrami, R. Kallosh, A. Linde and V. Vardanyan, Fortsch. Phys. 67, no.1-2, 1800075 (2019).
\bibitem{8}
 T. Brennan, F.Carta and C. Vafa, PoS TASI2017, 015 (2017).
\bibitem{9}
H. Murayama, M. Yamazaki and T. Yanagida, JHEP 12, 032 (2018).
\bibitem{10}
Eran. Palti, JHEP 2017, 34 (2017).
\bibitem{11}
Karta, Kooner, Susha Parameswaran and Ivonne Zavala, Phys. Lett. B 759, 402409 (2016).
\bibitem{12}
Miguel Montero, Gary Shiu and Pablo Soler, JHEB 2016, 159 (2016).
\bibitem{13}
Prashant Saraswat, Phys. Rev. D 95, 025013 (2017).
\bibitem{14}
Stefano Andriolo, Daniel Junghans, Toshifumi Noumi and Gary Shiu, Fortschritte der Physik 66, 1800020 (2018).
\bibitem{15}
Yu Akayama and Yasunori Nomura, Phys. Rev. D 92, 126006 (2015).
\bibitem{16}
Ben Heidenreich, Matthew Reece and Tom Rudelius, arxiv:1606.08437 (2016).
\bibitem{17}
Z. Yi and Y. Gong, arxiv:1811.01625 (2018).
\bibitem{18}
M. Artymowski and I. Ben-Dayan, JCAP 1905, 042 (2019).
\bibitem{19}
S. Brahma and M. Hossain, JHEP 2019, 6 (2019).
\bibitem{20}
David Andriot, Physics Letters B, 785:570{573 (2018).;
J. Sadeghi, S. Noori Gashti, and E.Naghd Mezerji. Phys. Dark Univ 30, 100626, doi: 10.1016/j.dark.2020.100626
\bibitem{21}
Gia Dvali and Cesar Gomez, Fortschritte der Physik, 67(1-2):1800092 (2019).;
J Sadeghi, E Naghd Mezerji and S Noori Gashti Modern Physics Letters A doi:10.1142/S0217732321500279 (2020).
\bibitem{22}
J. Sadeghi, S. Noori Gashti, Eur. Phys. J. C 81, 301 (2021).;
David Andriot and Christoph Roupec, Fortsch. Phys. 67(1-2):1800105 (2019).
\bibitem{23}
Sumit K Garg and Chethan Krishnan, arXiv:1807.05193 (2018).;
J Sadeghi, B Pourhassan, S. N Gashti, E. N Mezerji, and A Pasqua, arXiv:2108.01448 (2021).;
M Shokri, J Sadeghi, and S.N. Gashti, arXiv:2107.04756 (2021).
\bibitem{24}
Hirosi Ooguri, Eran Palti, Gary Shiu, and Cumrun Vafa, Physics Letters B, 788:180184 (2019).
\bibitem{25}
Tom Rudelius, arXiv:1905.05198, (2019).
\bibitem{26}
Jerome Martin and Robert H Brandenberger, Physical Review D, 63(12):123501, (2001).
\bibitem{27}
Cliford Cheung and Grant, Remmen, JHEP12, 087 (2014).
\bibitem{28}
Brando Bellazzini, Matthew Lewandowski and Javi Serra, Phys. Rev. Lett.123 25, 251103 (2019).
\bibitem{29}
H. Ooguri, E. Palti, G. Shiu, and C. Vafa, Phys. Lett. B 788 180-184 (2019).
\bibitem{30}
G. Obied, H. Ooguri, L. Spodyneiko, and C. Vafa, arxiv:1806.08362 (2018).
\bibitem{31}
S. Tsujikawa, Class. Quant. Grav. 30, 214003 (2013).
\bibitem{32}
H. Fukuda, R. Saito, S. Shirai, and M. Yamazaki,  Phys. Rev. D 99 8, 083520 (2019).
\bibitem{33}
S.-J. Wang, Phys. Rev. D 99 2, 023529 (2019).
\bibitem{34}
S. Das, Phys. Rev. D 99 6, 063514 (2019).
\bibitem{35}
I. Antoniadis, Y. Chen, and G. K. Leontaris, Int. J. Mod. Phys. A 34 08, 1950042 (2019).
\bibitem{36}
A. Ashoorioon, Phys. Lett. B 790 568-573 (2019).
\bibitem{37}
M. Motaharfar, V. Kamali, and R. O. Ramos, Phys. Rev. D 99  6, 063513 (2019).
\bibitem{38}
 S. D. Odintsov and V. K. Oikonomou, EPL 126 2, 20002 (2019).
\bibitem{39}
K. Dimopoulos, Phys. Rev. D 98 12, 123516 (2018).
\bibitem{40}
M. Kawasaki and V. Takhistov, Phys. Rev. D 98  12, 123514 (2018).
\bibitem{41}
K. Hamaguchi, M. Ibe, and T. Moroi, JHEP 12 023 (2018).
\bibitem{42}
C.-M. Lin, K.-W. Ng, and K. Cheung, Phys. Rev. D 100 2, 023545 (2019).
\bibitem{43}
L. Anguelova, E. M. Babalic, and C. I. Lazaroiu, JHEP 04  148 (2019).
\bibitem{44}
J. Ellis, B. Nagaraj, D. V. Nanopoulos, and K. A. Olive, JHEP 11  110 (2018).
\bibitem{45}
J. Halverson and F. Ruehle, (2018), Phys. Rev. D 99  4, 046015 (2019).
\bibitem{46}
 R. Brandenberger, R. R. Cuzinatto, J. Frohlich, and R. Namba, JCAP 02  043 (2019).
\bibitem{47}
I. Bena, E. Dudas, M. Gra˜na, and S. Lust, Fortsch. Phys. 67 1-2 (2019) , 1800100, Fortsch. Phys. 180010(2018).
\bibitem{48}
J. Moritz, A. Retolaza, and A. Westphal, Fortsch. Phys. 67 1-2, 1800098 (2019).
\bibitem{49}
L. Visinelli and S. Vagnozzi, Phys. Rev. D 99 6, 063517 (2019).
\bibitem{50}
 G. D’Amico, N. Kaloper, and A. Lawrence, Phys. Rev. D 100  10, 103504 (2019).
\bibitem{51}
C. Han, S. Pi, and M. Sasaki, Phys. Lett. B 791  314-318 (2019).
\bibitem{52}
 R. H. Brandenberger,  arXiv:1809.04926 (2018).
\bibitem{53}
Martin. Jerome, Ringeval. Christophe and Vennin. Vincent, Phys. Dark Univ. 5-6, 75--235 (2014).
\bibitem{54}
F. L. Bezrukov and M. Shaposhnikov, Phys. Lett. B659, 703 (2008).
\bibitem{55}
F. Bezrukov and M. Shaposhnikov, JHEP 07, 089 (2009).
\bibitem{56}
Y. Hamada, H. Kawai, K.-y. Oda, and S. C. Park, Phys. Rev. Lett. 112, 241301 (2014).
\bibitem{57}
Y. Hamada, H. Kawai, K.-y. Oda, and S. C. Park, Phys. Rev. D91, 053008 (2015).
\bibitem{58}
F. Bezrukov, J. Rubio, and M. Shaposhnikov, Phys. Rev. D92, 083512 (2015).
\bibitem{59}
F. Bezrukov and M. Shaposhnikov, Phys. Lett. B734, 249 (2014).
\bibitem{60}
Y. Akrami et al. (Planck), arXiv:1807.06211 [astro-ph.CO], (2018).
\bibitem{61}
N. Aghanim et al. (Planck), arXiv:1807.06209 [astro-ph.CO], (2018).
\bibitem{62}
P. A. R. Ade et al. (Planck), Astron. Astrophys. 594, A13 (2016).
\bibitem{63}
P. A. R. Ade et al. (BICEP2, Keck Array), Phys. Rev. Lett. 121, 221301(2018).
\bibitem{64}
E. Pajer, JCAP 0804  031,(2008).
\bibitem{65}
S. Kuperstein, JHEP 0503 014(2005).
\bibitem{66}
S. Kachru, R. Kallosh, A. D. Linde, and S. P. Trivedi, Phys.Rev. D68 046005(2003).
\bibitem{67}
Hao Geng, Physics Letters B,805, 10 (2020).

}
\end{thebibliography}
\end{document}